\documentclass[aps,prl,preprint,superscriptaddress]{revtex4}
\usepackage{graphicx}
\usepackage{verbatim}
\usepackage{ulem}
\usepackage{color}

\begin{document}

\title{Disentangling a quantum antiferromagnet with resonant inelastic X-ray scattering}

\author{L. Andrew Wray}
\affiliation{Advanced Light Source, Lawrence Berkeley National Laboratory, Berkeley, CA 94720, USA}
\author{Ignace Jarrige}
\affiliation{SPring-8, Japan Atomic Energy Agency, Hyogo 679-5148, Japan}
\author{Kazuhiko Ikeuchi}
\affiliation{SPring-8, Japan Atomic Energy Agency, Hyogo 679-5148, Japan}
\affiliation{Comprehensive Research Organization for Science and Society (CROSS), Ibaraki, 319-1106, Japan}
\author{Kenji Ishii}
\affiliation{SPring-8, Japan Atomic Energy Agency, Hyogo 679-5148, Japan}
\author{Yuri Shvyd'ko}
\affiliation{Advanced Photon Source, Argonne National Laboratory, Argonne, IL 60439, USA}
\author{Yuqi Xia}
\author{M. Zahid Hasan}
\affiliation{Department of Physics, Joseph Henry Laboratories, Princeton University, Princeton, NJ 08544, USA}
\author{Charles Mathy}
\affiliation{ITAMP, Harvard-Smithsonian Center for Astrophysics, Cambridge, MA 02138, USA}
\author{Hiroshi Eisaki}
\affiliation{Nanoelectronic Research Institute, National Institute of Advanced Industrial Science and Technology, Tsukuba, 305-8568, Japan}
\author{Jinsheng Wen}
\affiliation{Physics Department, University of California, Berkeley, CA 94720, USA}
\affiliation{Condensed Matter Physics and Materials Science Department, Brookhaven National
Laboratory, Upton, NY 11973, USA}
\author{Zhijun Xu}
\author{Genda Gu}
\affiliation{Condensed Matter Physics and Materials Science Department, Brookhaven National
Laboratory, Upton, NY 11973, USA}
\author{Zahid Hussain}
\author{Yi-De Chuang}
\affiliation{Advanced Light Source, Lawrence Berkeley National Laboratory, Berkeley, CA 94720, USA}

\begin{abstract}

Low-dimensional copper oxide lattices naturally manifest electronic states with strong short range quantum entanglement, which are known to lead to remarkable emergent material properties. However the nanometer scale many-body wavefunction is challenging to measure or manipulate in a simple way. In this study, X-ray induced $dd$ electronic transitions are used to suppress spin entanglement across a single lattice site in the spin-1/2 antiferromagnetic chain compound SrCuO$_2$, revealing a class of cuprate magnetic excitations that result from breaking the spin chain. These measurements are the first to employ two closely spaced X-ray resonances of copper (M$_2$ and M$_3$) as a form of natural 2-slit interferometer to distinguish between different types of electronic transition and resolve how they influence the dynamics of nearby spin-entangled electrons.

\end{abstract}

\date{\today}

\maketitle

Low-dimensional copper oxide lattices naturally manifest strong short range electronic interactions that cause remarkable emergent properties, such as high critical temperature superconductivity (high T$_C$) and the appearance of exotic particles defined by the collective dynamics of multiple electrons \cite{Mott,Allen,AndersonStrC,NagaosaStrC,OrensteinStrC,CKimSpinon}. Directly observing the microscopic quantum entanglement of these materials requires a specially selected probe that can single out an individual atom and perform precise manipulations on a femtosecond time scale to measure the many-body collective excitations that occur when the quantum entanglement of neighboring electrons is broken or altered. To achieve this goal, we have used resonant inelastic X-ray scattering (RIXS) to generate electronic transitions confined to a single copper lattice site that disconnect magnetic coupling in the antiferromagnetic spin $\frac{1}{2}$ chains of the quasi-one dimensional cuprate SrCuO$_2$. Our measurements make use of the closely spaced M$_2$ and M$_3$ resonances of copper as an analogue to the 2-slit interferometer, and use the pattern of scattering interference in conjunction with dramatically improved energy resolution to explore a previously unobserved class of cuprate magnetic excitations that result from breaking the spin chain. Patterns of quantum interference between M$_2$ and M$_3$ resonance channels are shown to unambiguously reveal the atomic orbital component and scattering phase difference of low energy excitations, critical pieces of information that would otherwise require ultrafast coherent measurements of a type not yet developed. Our results demonstrate that similar measurements have the potential to reveal quantum interference and magnetic dynamics in
quantum materials based on Ni, Co, Mn and Fe.

The cuprate SrCuO$_2$ (SCO) has been of interest in recent years as a model realization of a one dimensional (1D) spin $\frac{1}{2}$ antiferromagnet \cite{CKimSpinon,ZXOrbitals,HasanSCO_KRIXS, ZaliznyakNeutron,UchidaSusceptibility,TohyamaMade}, and as part of a class of quasi-1D strongly correlated materials manifesting extremely large nonlinear optical effects \cite{KishidaONL}. The copper oxide lattice is composed of chains of corner sharing copper oxygen plaquettes (Fig. 1C), with a $\frac{1}{2}$ spin moment in the singly occupied  3d$_{x^2-y^2}$ orbital on each Cu atom. The chains are formed in pairs that are weakly connected through magnetically frustrated edge-sharing plaquette boundaries. Spin interactions along the chain axis are unusually strong for a cuprate, with a Heisenberg coupling term of J$_H$=0.23eV measured from optical, neutron and photoemission spectroscopies \cite{CKimSpinon, ZaliznyakNeutron, PopovicOptical}. Like many High-T$_C$ superconductors, the crystal structure of SCO lacks apical oxygens above and below the Cu site, leading to a large crystal field splitting effect differentiating the energies of Cu 3d orbitals.

The orbital arrangement of electrons surrounding a copper atom can be changed by inducing a crystal field excitation. To do this, the energy of incident X-rays is set near the copper M$_2$ and M$_3$ resonances, corresponding to the excitation of electrons from spin-orbit coupled Cu 3p$_{1/2}$ and 3p$_{3/2}$ core levels to the 3d$_{x^2-y^2}$ orbital. Crystal field $dd$ excitations form when an electron from a different 3d orbital (e.g. 3d$_{xy}$ in the Fig. 1B diagram) transitions to the vacated 3p core level, leaving the hole in a new d-orbital as labeled on the RIXS spectrum in Fig. 1A. The electron added to the 3d$_{x^2-y^2}$ orbital is decoupled from the upper Hubbard band due to the attraction of the valence hole \cite{SalaLedge,Jeroen1Dfluct}, causing the excitations to be essentially immobile across the Brillouin zone (Fig. 1A, inset). Our high resolution data reveal that the onset edge of the lowest energy excitation closely resembles a 150meV Lorentzian (Fig. 1A), suggesting that the excited state has a long lifetime of $\tau_{xy}$$\sim$4.3fs ($\hbar$/150meV=4.3fs). Due to the strong spin interactions of SCO, these $dd$ excitations create an unusual situation in which the electronic configuration surrounding a single copper atom is reshaped into a metastable state that recovers more slowly than the characteristic intersite propagation of magnetic collective modes \cite{CKimSpinon}.

Since RIXS spectroscopy at transition metal M$_2$ and M$_3$ edges has not been extensively utilized, exploratory measurements of the $dd$ excitations across a wide range of incident photon energies are presented in Fig. 2A (bottom). Summing the RIXS intensity within 1-3.5eV excitation energy results in a curve representing the total probability that a $dd$ mode of any symmetry will be excited, as a function of incident energy (Fig. 2A, top). Two humps centered on the M$_2$ and M$_3$ resonance energies can be fitted using closely spaced Lorentzian functions with a 1:2 intensity ratio as expected from the relative core level degeneracies. In contrast, separately tracing the resonant scattering intensity of individual $dd$ modes in Fig. 2E reveals distinctively different patterns. The 3d$_{3z^2-r^2}$ $dd$ excitation primarily resonates in a single peak around h$\nu$=76eV, while the 3d$_{xy}$ excitation resonates over a broader energy range with two distinct peaks approximately 3eV apart, more widely spaced than the actual energy gap between 3p$_{1/2}$ and 3p$_{3/2}$ core levels.

The considerable overlap between M$_2$ and M$_3$ resonance suggests that the $dd$ modes are excited via intermediate states in which the core hole occupies a quantum superposition of the 3p$_{1/2}$ and 3p$_{3/2}$ levels. Within the widely successful mean field ``direct RIXS" scattering model \cite{SalaLedge,PattheyML,NiORIXS,MEcalc,AmentRIXSReview}, overlapping the resonant scattering signal from closely spaced core levels will produce an effect similar to conventional two-slit interference. The interference pattern is not caused by wavefronts intersecting in space, but rather by the phase of quantum paths through 3p$_{1/2}$ and 3p$_{3/2}$ states in the energy/time domain (Fig. 2D) as presented in the following scattering equation:

\begin{eqnarray}
R(h\nu,dd) \propto \left|\frac{A_{M2,dd}e^{i\theta_{M2,dd}}}{h\nu-E_{M2}+i\Gamma/2} + \frac{A_{M3,dd}e^{i\theta_{M3,dd}}}{h\nu-E_{M3}+i\Gamma/2}\right|^2
\end{eqnarray}

Resonant scattering intensity of a selected excitation mode (R(h$\nu$,$dd$)) is determined based on different real-valued quantum amplitudes and phases for scattering from M$_2$ and M$_3$ (e.g. A$_{M2,xy}$, $\theta_{M2,xy}$ for the 3d$_{xy}$ mode), as well as the resonance energies (E$_{M2}$=74.8eV,E$_{M3}$=77.55eV) and inverse resonance lifetime ($\Gamma$=2.5$\pm$0.1eV=$\hbar$/(0.26$\pm$0.3fs)). A wide range of information about the M-edge scattering process can be determined from these fitting parameters which will be discussed further in the Supplementary Online Material (SOM), including the M$_3$-resonance per-atom scattering cross section which we find to be larger than for the higher energy L$_{3}$ resonance by a factor of roughly two. As shown in Fig. 2E, the quantum interference effect described by Equation (1) nicely reproduces the squeezing together and spreading apart of resonant intensity observed in the data.

The distinctive forms of these resonance patterns occur because the spin and orbital
transitions involved in creating a $dd$ excitation through the M$_2$ and M$_3$ intermediate states
differ by a phase that shifts the constant phase contours of light scattered from M$_2$ and M$_3$
along the time axis (arrows in Fig. 2D). Using the resonance energetics parameters fitted
from the data, one can predict the full resonance profile in the direct RIXS process \cite{SalaLedge} and
obtain scattering phases of $\theta_{M2}$-$\theta_{M3}$$\sim$$\pi$ for 3d$_{3z^2-r^2}$ excitations and $\sim$$0$ for 3d$_{xy}$ in agreement with those fitted in Fig. 2E. Adopting a lowest order approximation that the $dd$ excitations
are Lorentzians along the energy loss axis, the resulting numerically derived spectrum is a
good qualitative match for the experimental data, as can be seen in Fig. 2A where the two are
overlaid. Previous studies have noted the existence of these classes of quantum interference, but have not established or made use of them experimentally by tracing the incident energy dependence of individual excitation modes \cite{Kuiper,NiORIXS}. In these measurements we experimentally observe that the phase shifts picked up during resonant scattering provide a new element of spectroscopic information to determine the orbital and spin symmetry of RIXS features which can otherwise be a subject of controversy and model-dependent speculation \cite{SalaLedge,Falck,GhirEarlyRIXS,Lorenzana,Kuiper}.

In RIXS experiments performed with resonance at more tightly bound core levels (e.g. K-edge), higher order ``indirect RIXS" processes involving a large continuum of highly complex intermediate resonance states are known to create patterns of incident energy dependence \cite{HasanSCO_KRIXS,Lu3rdOrder,VDBultrashort,Abbamonte3rdOrder,VDBphonons,DevereauxNew,AmentRIXSReview} that are not found in M-resonance data \cite{PattheyML,NiORIXS}. In indirect RIXS processes, the dynamics excited by a scattered X-ray are biased by the choice of incident photon energy and can become difficult to relate to desired correlation functions. These mechanisms cannot be responsible for the resonance patterns observed in our M-resonance data (Fig. 2E), as the characteristic indirect RIXS charge transfer excitations of SCO are not present (see Fig. 4C and Ref. \cite{HasanSCO_KRIXS}). The phase interferometric resonance patterns identified in our data are much simpler, caused by just two intermediate resonance states that have the same configuration of low energy electrons regardless of the incident energy used for measurement. Because of this, the dynamics induced by each separate $dd$ excitation are almost entirely independent of the incident photon energy, and the quantum interference pattern can be used as a tool to selectively enhance a specific excitation mode that one wishes to measure without compromising the excitation line shape.

Of the excitations seen in our data, the 3d$_{xy}$ $dd$ mode has the longest lifetime and is expected to cause the greatest disturbance to the many body electronic structure of surrounding lattice sites. Calculating the resonant phase interference pattern reveals that the 3d$_{xy}$ excitation can be observed with maximal intensity in the YZ scattering plane and is uniquely enhanced relative to other features at X-ray energies of h$\nu$$\sim$74.5eV, just below the M$_3$ resonance (Fig. 3A). At this photon energy, we rotate the sample azimuthal angle by 90$^o$ to selectively suppress the 3d$_{xy}$ excitation relative to other features, so that its line shape is observed `in isolation' by taking the difference between measurements in the YZ and XY scattering planes (Fig. 3C). The spectra measured in these two geometries can be directly compared because M-resonance RIXS involves very limited momentum transfer from low energy photons (Q$\sim$0) and a nearly featureless absorption cross-section dominated by a flat non-resonant background (see SOM). The isolated 3d$_{xy}$ $dd$ mode has an asymmetric line shape, also somewhat visible in our lower resolution K-pre-edge data (Fig. 3C, inset), with a sharp rising edge and shallow trailing edge that cannot be attributed to the charge and spin features known from earlier RIXS studies \cite{SalaLedge,NiORIXS}.

The line shape of $dd$ excitations encodes the rich many body dynamics brought about by disturbing the complex electronic structure of a cuprate. However, limited resolution has previously made it impossible to distinguish $dd$ mode profiles from the Lorentzian contours predicted in single atom mean field models \cite{SalaLedge,NiORIXS}. To understand where the non-Lorentzian asymmetry in our 3d$_{xy}$ data comes from, we consider the inherent quantum fluctuations of a spin $\frac{1}{2}$ Heisenberg antiferromagnet, which are particularly strong in 1D materials \cite{Jeroen1Dfluct} such as SCO and are proposed to provide the glue for cuprate superconductivity in quasi-2D lattices. Introducing a localized 3d$_{xy}$ $dd$ excitation to an antiferromagnetic chain in SCO breaks spin interactions by doubly-occupying the Cu orbitals that interact with neighboring spins (3d$_{x^2-y^2}$ and 3d$_{3z^2-r^2}$), causing the final state to be composed of two broken spin chains that terminate to the right and left of the excitation site. An exact diagonalization simulation of spin dynamics after breaking the SCO spin chain is shown in Fig. 4A, using the spin interaction term of J$_H$=0.23eV \cite{CKimSpinon, ZaliznyakNeutron, PopovicOptical, TohyamaMade}. At the instant when the excitation is created, magnetic moments close to the excitation site are correlated in a short range antiferromagnetic pattern that does not change significantly over the core hole lifetime ($\delta$t$<$0.3fs). As time evolves towards the full lifetime of the 3d$_{xy}$ $dd$ excitation ($\tau$$\sim$4.3fs), adjacent spins lose most of their correlation with the original spin of the transition site and the magnetic configuration up to 6 lattice sites away (2.4nm) changes significantly. Excited magnetic harmonics of the simulated broken chain are therefore seen to have significant dynamics over the lifetime of a $dd$ excitation, making them important in determining the spectrum observed by RIXS. Calculating the 3dxy excitation spectrum within an ideal J$_H$=0.23eV spin chain reveals that it closely matches the experimental line shape due to a high energy tail of spin
excitations (Fig. 4B).

To evaluate the importance of spin excitations in other features as well, the entire spectrum is fitted using identical parameters with spin dynamics turned on and off in Fig. 4C(right). When spin dynamics are discarded by setting J$_H$=0, or equivalently by treating the system as a classical Ising antiferromagnet, qualitative deviations are observed for all three $dd$ features. This clearly shows that excited magnetic collective modes of the broken spin chain are expected to contribute significantly to the $dd$ excitation spectrum. These spin modes occur due to quantum spin fluctuations and are not a feature of mean field spin models, unlike the local spin flip excitations that can effect the spectrum of the 3d$_{3z^2-r^2}$ mode \cite{NiORIXS,MEcalc,SalaLedge}. Similar line shape effects have been observed previously related to attosecond timescale charge relaxation near a perturbed site in simple molecules \cite{AbbamonteAtto}, but magnetic
dynamics in a cuprate occur on a comparatively slow femtosecond timescale and would be challenging to observe without high energy resolution achieved at the M$_2$- and M$_3$- resonance channels.

Another fundamental effect governing the line shape is the lifetime of RIXS charge excitations, which has been implicated as causing charge excitation profiles to be consistently broader than $\delta$E$\gtrsim$150meV in late transition metal oxides \cite{SalaLedge,GhirEarlyRIXS,Kuiper,PattheyML,NiORIXS}. Strikingly, our high resolution data reveal that creating RIXS excitations on isolated spectator ions such as Nd in nickelate Nd$_{2-x}$Sr$_x$NiO$_{4+\delta}$ can lead to a much sharper line shape ($\Delta$E=110$\pm$10meV in Fig. 4B, inset). Contrasting these two cases suggests that the intrinsic width observed by RIXS may be increased by electronic interactions within the transition metal oxide lattice. To consider dynamical modes that $dd$ excitations may interact with, the M$_3$-resonance $dd$ excitation spectrum in Fig. 4C is superimposed with measurements of excitations across the copper-oxygen charge transfer gap determined by K-edge RIXS and optical ellipsometry. The leading edge of the 3dxy $dd$ excitation does not overlap with a large density of charge transfer excitations, and has a sharper line shape than any other feature in the spectrum. As the energies of the $dd$ excitations increase beyond the charge transfer gap energy, a broadening trend is observed. The 3d$_{3z^2-r^2}$ excitation, which is consistently the broadest $dd$ mode
across many classes of cuprates \cite{SalaLedge}, has the greatest quantum overlap with charge transfer
excitations and can easily decay into a delocalized charge transfer collective mode.

The overall picture that emerges is one in which $dd$ excitations suppress spin interactions on a single lattice site for long enough to change the local correlation and singlet-like entanglement between spins \cite{AndersonStrC}, and then decay due to dynamical interactions with the surrounding crystal. The spin excitations triggered by a $dd$ mode occur solely due to non-classical quantum spin fluctuations and are thus expected to become weaker in more classical spin systems such as 2D cuprates and materials with larger local magnetic moment \cite{Jeroen1Dfluct}. Some
of the intensity in the modeled RIXS line shape can be attributed to freely propagating double- and quadruple- (etc.) spinon excitations, however other changes in the spin state are localized near the $dd$ excitation site and do not necessarily resemble free spinons.

These spin modes induced by $dd$ crystal field excitations are a new kind of many body excitation that can be explored with high energy resolution and excitation tunability from quantum interference at the M- resonances, and demonstrate how X-rays can interact with a complex quantum system by creating an immobile $dd$ excitation `plug' that blocks interac tions through a single lattice site. Our study also reveals a general method for observing the phase component of RIXS excitations by using the closely spaced transition metal M$_2$ and M$_3$ channels as a natural interferometer, a measurement that would otherwise require coherent phase-resolving experimental techniques that have not yet been implemented. Looking ahead, our data and analyses demonstrate a number of distinct features and concrete advantages to M-resonance RIXS over the much better known L- and K- resonance RIXS spectroscopies that will be important for future measurements, including an enhanced per-atom scattering cross section, strong quantum interference, simpler scattering matrix elements and a cleaner spectrum with fewer high-order excitation features from indirect RIXS.

\textbf{Methods summary:}

High resolution M-edge measurements were performed at the BL4.0.3 (MERLIN) RIXS endstation (MERIXS) at the Advanced Light Source (ALS), Lawrence Berkeley National Laboratory. The data were recorded by a VLS x-ray emission spectrograph equipped with a commercial Andor CCD detector cooled to -75$^o$C. Large single crystal samples were cleaved in situ and measured at a pressure of 3$\times$10$^{-10}$ Torr at room temperature. All data were obtained within 36 hours of cleavage to minimize possible aging of the cleaved surface. The resolution-limited peak width at half maximum of the elastic line is better than $\delta$E$\lesssim$35$\pm$2 meV for all M-edge data sets. High resolution measurements, with energy resolution of $\sim$90 meV, at the copper K-pre-edge were performed using MERIX spectrometer at the Advanced Photon Source IXS 30-ID beamline, and more comprehensive measurements including mapping of the K-edge dipole excitation spectrum were performed at SPring-8 BL11-XU. Numerical simulations and matrix elements relevant to RIXS scattering are discussed in the online SOM.

\textbf{Acknowledgements:}

We are grateful for fruitful discussions with A. Kotani, T. Tohyama, K. Tsutsui, K. Wohlfeld and W. Yang. The synchrotron radiation experiments at SPring-8 were performed under the Common Use Facility Programme of JAEA. Use of the Advanced Light Source and Advanced Photon Source was supported by the U.S. Department of Energy (DOE) Office of Science, Office of Basic Energy Sciences (contract no. DE-AC02-05CH11231 and DE-AC02-06CH11357).

\newpage

\begin{figure*}[t]
\includegraphics[width = 17cm]{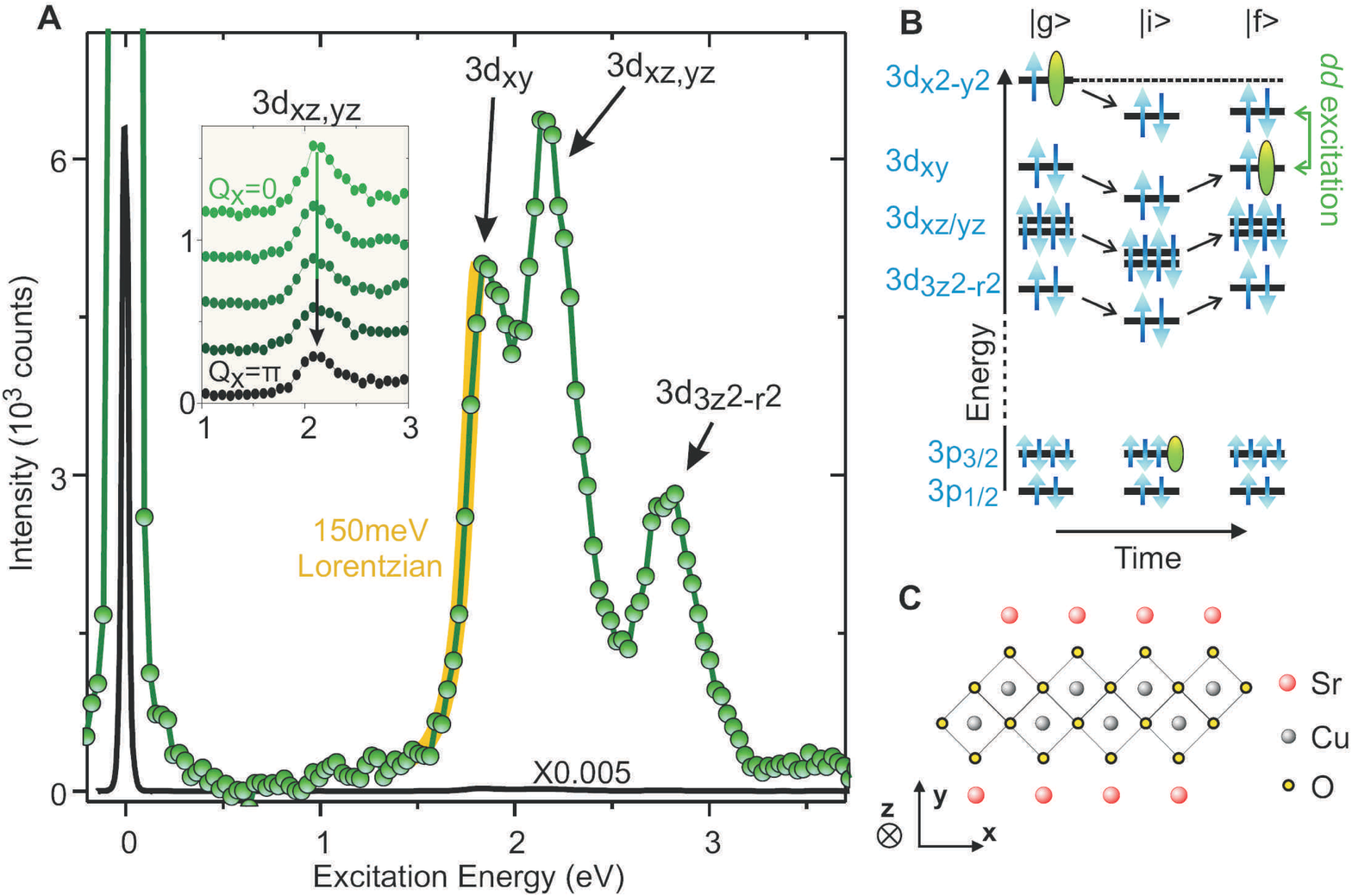}
\caption{{\bf{Massive single-atom excitations}}: (\textbf{A}) Excitations measured using copper M$_3$ resonant scattering on SrCuO$_2$ are primarily crystal field transitions (highly localized electron-hole pairs), labeled by the 3d orbital symmetry of the hole. Statistical error, similar to the data point size, is roughly constant across the spectrum due to dark noise of the detector. An inset shows momentum dispersion of the 3d$_{xz/yz}$ excitations measured at the K- pre-edge. (\textbf{B}) Ground, intermediate and final state ($|g\rangle$,$|i\rangle$,$|f\rangle$) electron configurations are shown for the direct RIXS scattering process by which $dd$ excitations are created using a 3p$_{3/2}$ (M$_3$) core level resonance. The energy of the 3d$_{x^2-y^2}$ orbital in the absence of a core or valence hole is indicated by a dashed line, showing that $dd$ excitations are decoupled from the upper Hubbard band. Intra-orbital interactions of the 3d$_{x^2-y^2}$ and 3d$_{xy}$ electrons are neglected for representational simplicity. (\textbf{C}) The crystal structure of copper-oxygen chains in SrCuO$_2$ is shown, with axis labels following the convention of two dimensional cuprates.}
\end{figure*}

\begin{figure*}[t]
\includegraphics[width = 15cm]{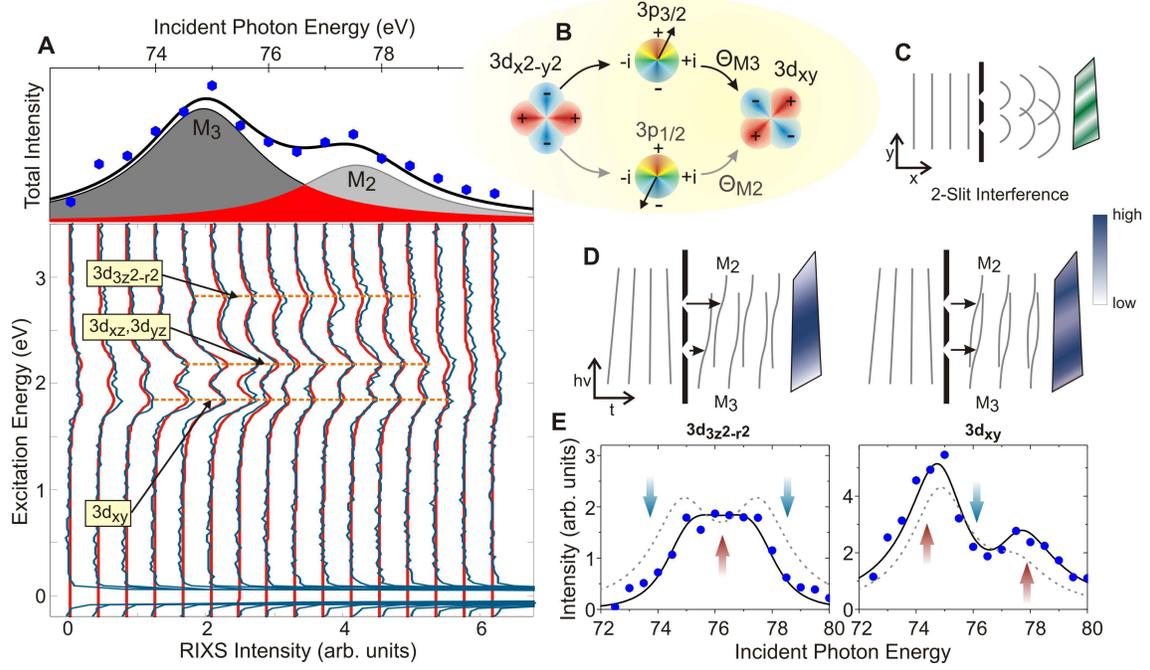}
\caption{{\bf{M$_2$-M$_3$ quantum interference}}: (\textbf{A}) (bottom,blue) Inelastic excitation spectra are shown with incident photons spanning the copper M$_2$ and M$_3$ resonance energies, and overlaid with (red) a theoretical prediction for the $dd$ modes labeled. (top) The summed inelastic intensity from each RIXS curve is compared with a two-Lorentzian fit. (\textbf{B}) The orbital transitions involved in creating a 3d$_{xy}$ excitation through M$_2$ and M$_3$ core hole resonances are shown. (\textbf{C}) The constant phase contours in a conventional 2-slit interference experiment. (\textbf{D}) The constant phase contours from scattering through M$_2$ and M$_3$ resonance are offset in time by a phase that depends on the final excitation symmetry, generating distinctive interference patterns for (left) 3d$_{3z^2-r^2}$ and (right) 3d$_{xy}$ $dd$ modes. (\textbf{E}) (circles) Incident energy dependence of the 3d$_{xy}$ and 3d$_{3z^2-r^2}$ modes is fitted from Equation (1) (solid line) with phase shifts $\theta_{M3,z^2}=\theta_{M2,z^2}+\pi$ and $\theta_{M3,xy}=\theta_{M2,xy}$. A dashed line shows the fit result with quantum interference disregarded, and arrows highlight the change from quantum interference.}
\end{figure*}

\begin{figure*}[t]
\includegraphics[width = 15cm]{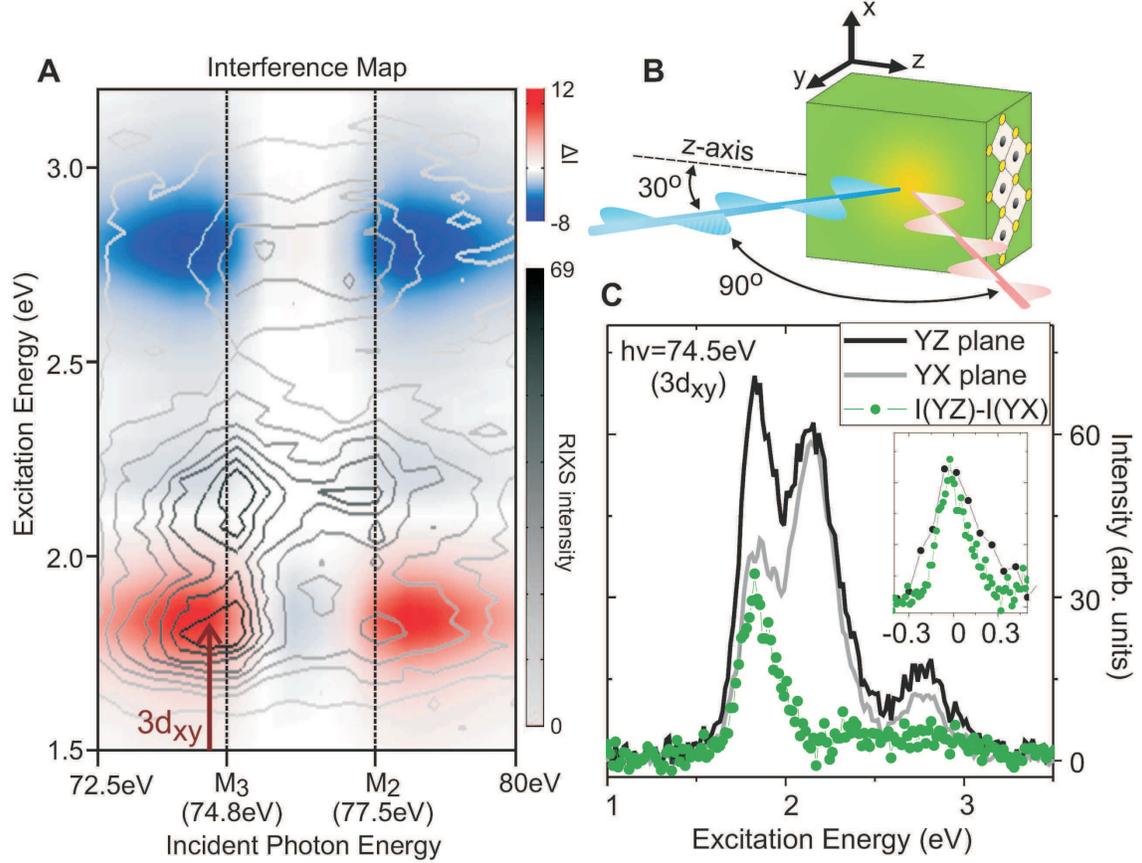}
\caption{{\bf{Quantum interference and the excitation line shape}}: (\textbf{A}) A contour plot of the SrCuO$_2$ M-resonance scattering data is overlaid with colors indicating the amount by which intensity is increased or decreased by quantum interference based on the orbital symmetries. An optimal energy for studying the 3d$_{xy}$ mode (h$\nu$=74.5eV) is indicated with a black arrow. (\textbf{B}) The scattering geometry is shown. (\textbf{C}) Changing the scattering plane by rotating the sample 90$^o$ around the azimuthal (y-) axis suppresses the intensity of the lowest energy $dd$ transition without greatly changing momentum because Q$_1$$\sim$Q$_2$$\sim$0. An inset contrasts (green, high resolution) the shape of a single $dd$ excitation revealed by M$_3$ geometrical contrast with (black) the shape of a $dd$ excitation observed at the K- pre-edge (resolution $\delta$E$\sim$100meV).}
\end{figure*}

\begin{figure*}[t]
\includegraphics[width = 16cm]{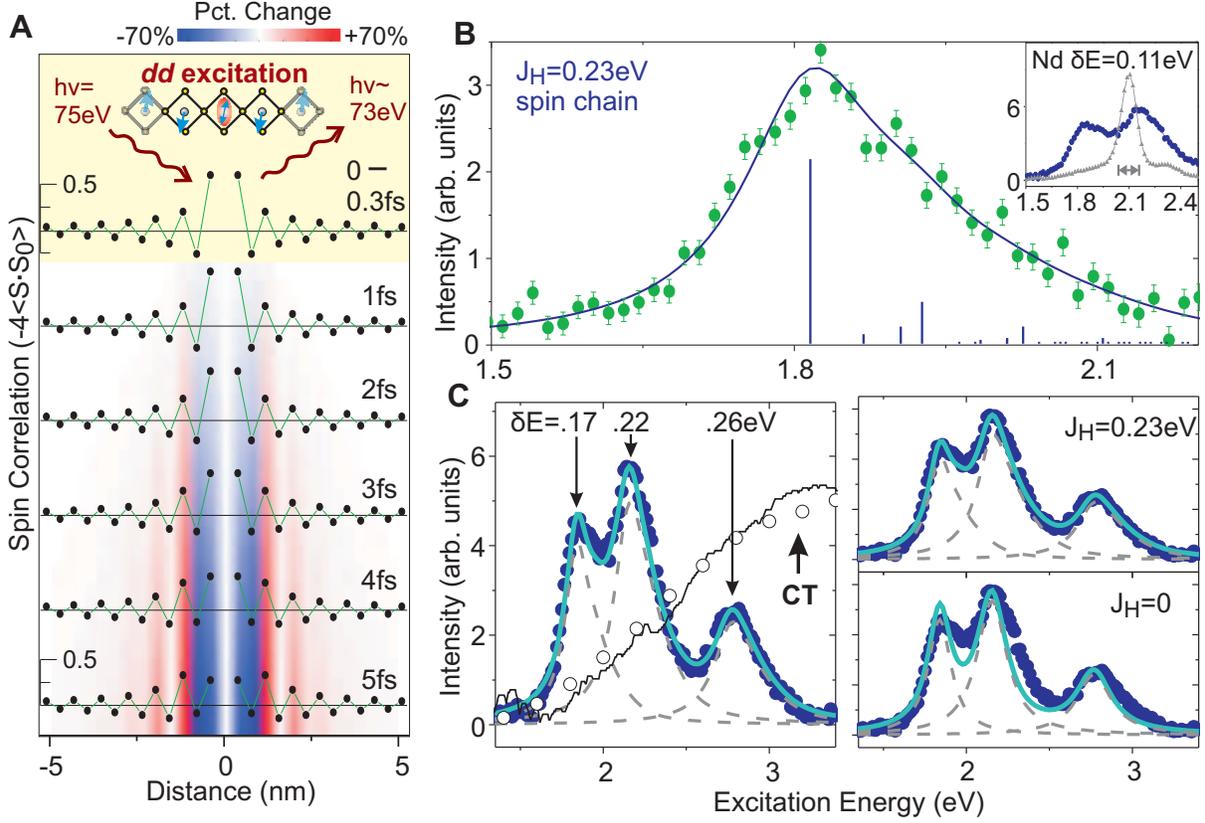}
\caption{{\bf{Excitation dynamics in a broken spin chain}}: (\textbf{A}) Time evolution of magnetic correlation with the original spin of the $dd$ excitation site is indicated by dots, and the percentage change relative to the t=0 magnetic structure is indicated with color shading. (\textbf{B}) The expected line shape of a single $dd$ excitation in the SrCuO$_2$ spin chain is obtained by diagonalizing a 20-site ring and (blue curve) applying $\delta$E=150meV lifetime broadening. (green circles) The experimental line shape of the 3d$_{xy}$ $dd$ excitation obtained by contrasting the RIXS signal in two different scattering geometries is shown with statistical error bars. An inset overlays the SrCuO$_2$ excitations with (silver) the $ff$ RIXS spectrum from Nd spectator atoms in Nd$_{1.67}$Sr$_{0.33}$NiO$_{4+\delta}$. (\textbf{C}) (left) The $dd$ transitions of SrCuO$_2$ measured with M$_3$ photons (h$\nu$=75eV) are fitted as excitations in a spin chain, using the inverse lifetimes ($\delta$E) indicated. Dipole induced charge transfer excitations measured from (open circles) K-edge RIXS and (black line) optical ellipsometry from Ref. \cite{PopovicOptical} are overlaid, revealing delocalized charge excitations that $dd$ crystal field modes can decay into. (right) Setting the spin interaction (J) to zero gives a prediction that neglects quantum spin fluctuations.}
\end{figure*}

\end{document}